\documentstyle[prb,aps,epsf,multicol]{revtex}
\begin{document}
\draft
\title{Curvature of Levels and Charge Stiffness of One-Dimensional Spinless Fermions}
\author{N. M. R. Peres$^1$, P. D. Sacramento$^2$, D. K. Campbell$^3$, and  J. M. P. 
Carmelo$^1$}
\address{$^1$Department of Physics, University of \'Evora,
Apartado 94, P-7001 \'Evora Codex, Portugal}
\address{$^2$Departamento de F\'{\i}sica and CFIF, Instituto Superior 
T\'ecnico, Av. Rovisco Pais, P-1096 Lisboa Codex, Portugal}
\address{$^3$ Department of Physics, University of Illinois at
Urbana-Champaign, 1110 West Green Street, Urbana, Illinois 61801}
\date{April 15, 1998}
\maketitle
\widetext
\begin{abstract}
Combining the Bethe Ansatz  with a {\it functional deviation expansion} and using an asymptotic expansion of the Bethe
Ansatz equations, we
compute the curvature of levels $D_n$ at any
filling for the one-dimensional lattice spinless fermion model. We use these results to study the finite temperature
charge stiffness $D(T)$. We find that
the curvature of the levels obeys, in general, the relation
$D_n=D_0+\delta D_n$, where $D_0$ is the zero-temperature charge stiffness and $\delta D_n$ arises from excitations.
Away from half filling and for the low-energy (gapless) eigenstates,
we find that the energy levels are, in general, flux 
dependent and, therefore, the system behaves as an ideal
conductor, with $D(T)$ finite. We show that
if gapped excitations are included the low-energy excitations feel an effective flux $\Phi^{eff}$ 
which is different
from what is usually expected.
At half filling,
we prove, in the strong interacting limit and to order
$1/V$ ($V$ is the nearest-neighbor Coulomb interaction), that 
the energy levels are flux independent. This leads to
a zero value for the curvature of levels $D_n$
and, as consequence, to $D(T)=0$,
proving an earlier conjecture of Zotos
and Prelov\v{s}ek. 
\end{abstract}
\begin{multicols}{2}

\section{Introduction}
\label{introdd}
The study of transport and response
properties in exactly solvable one-dimensional (1D) correlated
fermion systems, e. g. the spinless fermion, Hubbard, and $t-J$ models, has
long been an important field of theoretical research
\cite{Maldague77,Loh88,Schulz90,Stafford90,Fye91,Giamarchi92,Carmelo92,Horsch93,Castella95,Castella96,Zotos96,Antonio96,Naef97}.
This theoretical interest
has been supported by experimental studies that have found
both qualitative and quantitative agreement 
between thermodynamic, spectral, and
transport measurements in quasi-1D materials and the
corresponding properties calculated in 1D correlated electron models
\cite{Sacramento94,Eggert94,Wells95,Motoyama96,Kim96,Mori97}.

Recently, several comparative numerical and analytical studies
have explored the differences in the transport properties between
integrable and non-integrable 1D
models \cite{Castella95,Castella96,Zotos96,Antonio96,Naef97}.
Most of these studies have dealt with generalizations to finite-temperature
of Kohn's zero-temperature concepts and approach \cite{Kohn64}.
In reference \onlinecite{Zotos96}, the concepts of {\it ideal
insulator} and {\it ideal conductor} 
at {\it finite} temperatures were introduced.
These concepts refer to the 
temperature dependence of the real part of the optical
conductivity $\sigma_r(\omega,T)$, which is given by
\begin{equation}
\sigma_r(\omega,T)=2\pi D(T)\delta(\omega)+
\sigma_{reg}(\omega,T)\,,
\label{cond}
\end{equation}  
where we have taken $\hbar=e^2=1$, and $e$ being the electron charge.
The quantity $D(T)$ is called the charge stiffness and characterizes the
response of the
system to a static electric field, within linear response theory. 
According to Kohn's
zero-temperature criterion, the value of $D(0)$ can be used to 
distinguish between an
ideal insulator --$D(0)=0$ -- and an ideal conductor -- $D(0)\neq 0$. 
The quantity
$\sigma_{reg}(\omega,T)$ is called the regular part of the conductivity and
describes the absorption of light of finite frequency $\omega$ by the system.

At finite temperatures, the classification of a given system as an
{\it ideal conductor} or an {\it ideal insulator} involves the values of both
$D(T)$ and $\sigma_0(T)=\sigma_r(\omega \rightarrow 0,T)$ as follows: (i)
if $D(T)>0$, the system behaves as an {\it ideal conductor}; (ii) if $D(T)=0$ 
and
$\sigma_0(T)=0$ the system behaves as an {\it ideal insulator}; (iii) and,
finally, if $D(T)=0$ and $\sigma_0(T)>0$ the system 
behaves as a normal conductor.

Motivated by the results of
Refs. \onlinecite{Castella95,Castella96,Zotos96},
we compute in this article the curvature of levels $D_n$ and study
the effect of the temperature on the charge 
stiffness of the simplest 1D lattice
spinless fermion model, the Hamiltonian for which is

\begin{equation}
        \hat{H} = -\frac{J}{2} \sum_{\langle j,i \rangle}
        c_{j}^{\dag }c_{i}  +
        V\sum_{i}(\hat{n}_{i}-\frac 1{2})(\hat{n}_{i+1}-
        \frac 1{2})\,,
\label{spinless}
\end{equation}
where the spinless fermion operators $c_{i}^{\dag }, c_{i}$
obey the usual anti-commutation relations, $\langle i,j \rangle$ means
summation over nearest neighbors,  $\hat{n}_{i}=c_{i}^{\dag }c_{i}$
is the usual local number operator, $\frac{J}{2}$ 
is the hopping integral (normally called $t$ but here called
$\frac{J}{2}$ to stress the important exact correspondence between
this model and the (anisotropic) Heisenberg S=1/2 chain), and $V$ is the 
nearest-neighbor Coulomb repulsion.

The model\,(\ref{spinless}) is solvable by the Bethe Ansatz
(BA) \cite{Yang66,Cloizeaux62,Cloizeaux66} and has a metal-insulator
transition at half filling and zero temperature
(in this model half filling means a fermion for each {\it two} lattice sites,
i.e., fermionic density $1/2$). The
metal-insulator transition \cite{Cloizeaux66} occurs
when the first-neighbor Coulomb repulsion $V$ 
is greater than $J$. This is in contrast to the standard Hubbard model,
where the Mott-Hubbard transition
occurs at half filling for any non-zero value of the on-site
Coulomb repulsion \cite{Lieb68} (in the Hubbard model half filling
means one electron per lattice site, i.e., electron density $1$). 

The optical properties of the model (\ref{spinless}) can be
determined by studying the current operator $\hat{j}$, which is given
by
\begin{equation}
\hat{j}=i\frac{J}{2}\sum_{j}(c_{j}^{\dag }c_{j+1}-c_{j+1}^{\dag }c_{j})\,.
\label{cur}
\end{equation}
As in the case of the Hubbard chain, the commutator
of $\hat{j}$ with the Hamiltonian  (\ref{spinless})
is finite (i.e., non-zero) {\it operator} (i.e., not a $c-$number).
This implies that the real part of the optical conductivity has
finite-energy absorption, i.e., $\sigma_{reg}(\omega,T)\neq 0$.
The exact calculation of $\sigma_r(\omega,T)$ requires the full computation of
the Kubo formula. The changes of the energy eigenvalues
$E_n$ in response to an external flux $\phi$ piercing the ring
(formed by the 1D chain with periodic boundaries) can be determined
by solving Hamiltonian (\ref{spinless})
with the hopping integral modified by the usual Peierls phase factor:
$t\rightarrow te^{i\phi/N_a}$.

The charge stiffness $D(T)$ can be evaluated as a
thermodynamic quantity \cite{Castella95,Castella96,Zotos96} using
\begin{eqnarray}
        D(T)&=&\frac 1{N_a}\sum_n p_n
        D_n
        =\frac 1{2N_a}\left.
        \frac{d^2\,F}{d\phi^2}\right\vert_{\phi=0} 
\nonumber\\
&+&
        \frac 1{2TN_a}\sum_n p_n 
        \left ( j_n \right )^2\,,
\label{dt}
\end{eqnarray}
where $p_n=Z^{-1}e^{-\beta E_n}$ is the usual Boltzmann weight, $Z$ is the partition function, $2D_n=
d^2\,E_n/d\phi^2\vert_{\phi=0}$ is the curvature of the level $E_n$,  $j_n=-
d\,E_n/d\phi\vert_{\phi=0}$  is the mean value of the current operator in
the eigenstate of energy 
$E_n$,  $F$ is the free energy, and $N_a$ is the number of lattice sites.
The summation in Eq.\,(\ref{dt}) is 
not in general
simple to perform, and it must be computed for a given fermionic
density $n_f=N_f/N_a$, with $N_f$ the total number of fermions.

Many recent studies have provided crucial insight into various aspects
of this general problem. First, using the Mazur inequality, Zotos and
collaborators \cite{Naef97} derived some analytical inequalities
for $D(T)$, valid for $T\rightarrow\infty$.
Their results imply ideal conductivity away from half filling both
for the spinless fermion and Hubbard models but are inconclusive for the
half-filled band case. Second, for the Hubbard model, 
a recent analytical study by Fujimoto and
Kawakami \cite{Kawakami97}, using the thermodynamic BA \cite{Takahashi97},
confirmed the exponentially activated nature of the conductivity
$ D(T)\propto\sqrt(T)\exp(-\Delta_{MH}/T)$ ($\Delta_{MH}$ is the Mott-Hubbard
gap \cite{Lieb68}) at half filling, and 
ballistic transport of charge $D(T)=D_0-aT^2$
($a$ is a positive constant), for low temperatures and away from half filling. 
These results have been confirmed by Kirchner {\it et al.} 
combining both Monte Carlo and BA 
methods \cite{Kirchner98}. As a result of their study, 
these authors found numerical evidence that some of the 
conjectures of Refs.
\onlinecite{Castella95,Castella96,Zotos96} may not be correct.
Third, a recent study \cite{Tomaz98} of
the ``{\it kicked} $t-V$'' model has focused 
on the transition from integrability
to ``ergodicity,'' in the thermodynamic limit. This model is a modification of
the Hamiltonian (\ref{spinless}) in which the interaction term
$V$ is time dependent. Reference \onlinecite{Tomaz98}
related the infinite number of conservation 
laws \cite{Shastry86} that characterize
the integrable model (\ref{spinless}) to its {\it nonergodicity}, and
argued that the deviation from the {\it ergodic} behavior led to
the anomalous transport properties observed
in Refs. \onlinecite{Castella95,Castella96,Zotos96}. In particular,
 Ref. \onlinecite{Tomaz98} pointed out that {\it nonergodicity}
implies an infinite transport coefficient and, therefore, ideal conductivity
away from half filling. 
In the approach we develop in this paper, the 
infinite number of conservation laws manifest themselves
in the existence of {\it only} forward scattering among the exact many-body
excitations of the system (called pseudoparticles below) at all energy scales. 
Fourth, in the regime $-2t<V<2t$ (so that the interaction in Eq. (\ref{spinless}) 
is parametrized as $V=J\cos \lambda$) the behavior of $D(T)$ 
has been studied by Narozhny {\it et al.} using exact 
diagonalization and finite-size scaling
\cite{Narozhny98}. These authors found ideal conductivity 
behavior arising from a non-zero fraction of 
degenerate states in the thermodynamic limit.

It is worthwhile to remark that 
different authors associate ideal 
conductivity with different physical 
mechanisms -- conservation laws, nonergodicity, degeneracy of
the many-body energy states, 
forward scattering (see below) -- but at present the deep
relation (if any) among all of them remains unclear.

In this article, we focus on the regime ($ V > 2t$)
of the spinless fermion model
in which the metal-insulator transition can occur
and compute the needed mean values of the current operator in
any eigenstate for the model (\ref{spinless}), 
as well as the respective curvature of levels.
In the regime of the metal-insulator transition 
we introduce for later convenience the parametrization
$V=J\cosh \lambda$, with $\lambda >0$.

To carry out part of our study, we use the formalism of the
{\it pseudoparticle operator algebra}, which has recently been presented
in detail for the Hubbard model \cite{Nuno97b}. In brief,
the BA solution (discussed in more detail below) can be shown to refer to
an algebra of operators describing ``pseudoparticle''
excitations. The energy eigenstates are characterized by the pseudoparticle
momentum distributions, $N_{c,\gamma}(q)$. Here $c,\gamma $ with $\gamma
=0,1,2,3,...$ are the quantum numbers that label
the different pseudoparticle branches, with $N_{c,\gamma}(q)=0$ for $\gamma >0$
in the ground state. The BA gives the energy and other operator mean values
as functionals of the momentum distributions $N_{c,\gamma}(q)$.
Hence, as in references \cite{Carm91a,Carm91b}, we can employ a
{\it functional deviation expansion} (FDE) which refers to the momentum
{\it deviations}, 
$\delta N_{c,\gamma}(q) = N_{c,\gamma}(q)-N^0_{c,\gamma}(q)$,
where $N^0_{c,\gamma}(q)$ is the ground-state momentum
distribution. This leads to FDE expressions for the energy, 
current mean values,
and other mean values, exactly as in a Fermi liquid. In addition to the
central role played by
the FDE expansions, there are other similarities between 1D quantum liquids
and 3D Fermi liquids \cite{Baym}. These similarities
have justified the introduction of the concept of a ``Landau liquid''
\cite{Carm91a,Carm91b,Carm98}, 
which includes both the 3D Landau Fermi liquids\cite{Baym} and
the 1D Luttinger liquids \cite{Haldane81,Haldane91} as specific cases.
There is, however, an important difference between 
the Fermi liquid quasiparticles
and the pseudoparticles in 1D integrable models: namely, the collisions between
the quasiparticles are not, in general, of forward scattering type only,
whereas in the case of the pseudoparticles in integrable models, there is
only forward scattering. Moreover, the quasiparticles are not the
true many-body excitations of the quantum liquid. Therefore, it is not
to be expected that the transport properties behave in the same way for the
two types of quantum liquids.

The remainder of the paper is organized as follows. In Sec. \ref{lbethe}
we treat the BA equations in the thermodynamic limit
and compute the low- and high- energy excitation spectra,
and the corresponding energy gaps. We show that the elementary
excitations (the pseudoparticles) interact via forward
scattering interaction (the $f-$function) only, and we 
give its explicit form. We show that the energy eigenvalues
are fuctionals of the pseudoparticle
occupancies and depend on the excitation spectra and on the $f-$functions. Using the BA equations with a flux we obtain
general expression 
for $j_n$ and $D_n$, in the low-energy
Hilbert subspace. We develop the FDE, which leads to simpler expressions
for $j_n$ and $D_n$, for the low energy eigenstates. In Sec. \ref{asymptotics} we treat the BA equations in a finite-size
system and derive the curvature of levels $D_n$ as function of $V$ and $N_a$, in the limits
$V=\infty$ and $V\gg t$ ($1/V$ corrections), for all the eigenstates of the model. We use our results to discuss
the ballistic transport of charge at finite temperature
in this model.
In Sec. \ref{lconc} we discuss our results and present the conclusions.
\section{Bethe Ansatz Results}
\label{lbethe}

Our approach will be to use
the general BA solution \cite{Gaudin72} for the Hamiltonian (\ref{spinless})
with a flux $\phi$ through the ring \cite{Shastry90} to compute the
mean value of the current operator $j_n$ and the level curvature $D_n$
in a given eigenstate of energy $E_n$ using the formalism of Refs. 
\cite{Nuno97a,Nuno97b,Nuno97c}. Our results will follow directly from
the structure of the Hilbert space \cite{Gaudin72} 
of the model (\ref{spinless}).
In the following, we discuss this structure in a finite-sized system
and present a simple physical picture for this problem.

It is well known that the model (\ref{spinless}) can be mapped onto the
Heisenberg spin-one-half model by the Jordan-Wigner transformation.
For a given canonical ensemble there are states
with both $S_z=S$ and $S_z\ne S$, where $S$ is the total spin and
$S_z$ is its projection in the $z$ direction. We shall present the results
for those states that are characterized by $S_z=S$, but similar
results would be obtained for the states with $S_z\ne S$
(for example, the lowest energy state with $S=1$ and $S_z=0$
is degenerate, in the thermodynamic limit, with the ground state
characterized by $S=0$ and $S_z=0$).

The model (\ref{spinless})
has particle-hole symmetry and, therefore, without loss of generality
we study the case where the fermion density $n_f$ is less than $1/2$.
The general BA solution for the Hamiltonian (\ref{spinless}) is obtained
by combining the Jordan-Wigner transformation with the BA solution for the
Heisenberg spin-one-half model \cite{Gaudin72}. Our starting point is the set
of BA equations with (Orbach parametrization) 
twisted boundary conditions, which are \cite{Gaudin72,Shastry90}
\end{multicols}
\vspace{-0.6truecm}
\noindent\makebox[8.8truecm]{\hrulefill}
\widetext
\begin{equation}
        2\arctan\left ({A_{\gamma}
\tan{\frac {\Lambda_{c,\gamma}^j}{2}}}\right )=
        \frac{2\pi I_{c,\gamma}^j}{N_a}
+(\gamma+1)\frac {\phi}{N_a}+\sum_{\gamma'=0}\sum_{p=1}
        \frac {[\gamma\gamma'p]}
         {N_a}\sum_{j'}
        2\arctan \left ({A_p\tan{\frac {\Lambda_{c,\gamma}^j-
        \Lambda_{c,\gamma'}^{j'}}{2}}} \right )\,,
\label{bethe}
\end{equation}
\vspace{-0.4truecm}
\hspace*{\fill}\makebox[8.8truecm]{\hrulefill}
\begin{multicols}{2}
\narrowtext\noindent
where the numbers $\Lambda_{c,\gamma}^j$ are 
called rapidities, the label $c$ is
associated with charge, $A_{\gamma}=\coth[(\gamma+1) \lambda/2]$,
$A_p=\coth(p \lambda/2)$, $\phi$ is the twist angle, and the symbol
$[\gamma\gamma'p]$ is given by

\begin{eqnarray}
&&[\gamma\gamma'p] = 1, {\rm for }\hspace{.5cm} 
p=\vert \gamma-\gamma' \vert,\, 
\gamma+\gamma'+2 \nonumber\\
&&[\gamma\gamma'p] = 2, {\rm for } \hspace{.5cm}
p=\vert \gamma-\gamma' \vert+2, \vert \gamma-\gamma' 
\vert+4,\ldots\nonumber\\
&& \hspace{3.5cm}\ldots, \gamma+\gamma'\nonumber\\
&&[\gamma\gamma'p] = 0, \hspace{.5cm}{\rm otherwise}\,.
\end{eqnarray}
The energy $E_n$ of a given eigenstate is a function of the chosen set
of numbers $\{ I_{c,\gamma}^j \}_n$ and can be written as
\begin{equation}
        E_n=-J\sinh(\lambda)\sum_{\gamma}\sum_{\{ I_{c,\gamma}^j \}_n}
        \frac {\sinh[(\gamma+1) \lambda]}{\cosh[(1+\gamma )\lambda]-
\cos(\Lambda_{c,\gamma}^j)}\,.
\label{eig}
\end{equation}
The exact solution (\ref{bethe}) and (\ref{eig}) 
of the model (\ref{spinless}) shows
that all its eigenstates, for a given canonical ensemble, are characterized by
the quantum numbers $I_{c,\gamma}^j$, with $\gamma=0,1,2,3\ldots$. For a given
canonical ensemble, the number of fermions $N_f$ is related to the total number
$N_{c,\gamma}$ of occupied $I_{c,\gamma}^j$ quantum numbers by

\begin{equation}
        N_f=\sum_{\gamma=0}(\gamma+1)N_{c,\gamma}\,.
\label{sumrule}
\end{equation}
The numbers $I_{c,\gamma}^j$ can be 
positive or negative integers or half-odd integers depending 
on 
whether $N_a+N_{c,\gamma}$ are odd or 
even respectively ($N_a$ is taken to be even).
The eigenstates of the model are characterized by the set of numbers
$N_{c,\gamma}$ and by the configurations of the occupied quantum numbers
$I_{c,\gamma}^j$ among their possible values. 
The $N_{c,\gamma}$ numbers give the 
numbers of $c,\gamma$ pseudoparticles
and $q=2\pi I_{c,\gamma}^j/N_a$ is the 
momentum.
For a given canonical ensemble the $I_{c,\gamma}^j$ are distributed
in the interval

\begin{eqnarray}
        &&-\frac{Q_{\gamma}-1}{2}\le\,I_{c,\gamma}^j\,\le
        \frac{Q_{\gamma}-1}{2},\hspace{0.5cm}j=1,\ldots,N_{c,\gamma}\,,
\nonumber\\
        &&Q_{\gamma}=N_a-\sum_{\gamma'=0}M(\gamma,\gamma')N_{c,\gamma}\,,
\label{interval}
\end{eqnarray} 
and $M(\gamma,\gamma')=
2min(\gamma+1,\gamma'+1)$ for $\gamma \neq \gamma'$, and 
$M(\gamma,\gamma')=2\gamma+1$ for $\gamma = \gamma'$. 
The $I_{c,0}$ numbers describe the low energy excitations
while the $I_{c,\gamma>0}$ numbers 
are related to the BA string solutions and describe the gapped
excitations \cite{Gaudin72}.

\subsection{Bands, gaps, and $\lowercase{f}$-functions: $\phi=0$ results}

The numbers $N_{c,\gamma}$ and the
momentum $q$ introduced above 
have a simple physical interpretation. The index $\gamma$ labels a set of 
energy bands which have their (pseudo-) Brillouin-zones and (pseudo-)
Fermi points controlled by the relations (\ref{sumrule}) and (\ref{interval}). 
All these bands are separated from each other by energy gaps. These gaps, 
relative to the lower energy band $c,0$, are given by
\begin{equation} \Delta_{\gamma}=\varepsilon_{c,\gamma>0}(0)-
(\gamma+1)\varepsilon_{c,0}(q_{Fc,0})\,,
\label{gap}
\end{equation}
where $\varepsilon_{c,\gamma}(q)$ 
is the energy dispersion of the pseudoparticle
and $q_{Fc,\gamma}=2\pi I_{c,\gamma}^{max}/N_a$, with $I_{c,\gamma}^{max}$
being the largest quantum number $I_{c,\gamma}^j$, in the compact symmetric
distribution of these numbers around zero. For $n_f=1/2$, 
we have $q_{Fc,0}=\pi/2$
and $q_{Fc,\gamma >0}=0$ in the ground state. 
The dispersion relations for the $c,\gamma$ pseudoparticles 
at half filling are 
\begin{eqnarray}
\varepsilon_{c,0}(q)&=&-J\frac{\sinh \lambda}{2}-J
\frac{{\rm K'}\sinh\lambda}{\lambda}\sqrt{1-
k^2\sin^2(q)}\nonumber\\
\varepsilon_{c,\gamma}(0)&=&-J\sinh\lambda\,,
\label{dispsf}
\end{eqnarray}
and their general equations are given in the Appendix.
The parameter K$'$ denotes the 
complete elliptic integral, which 
is a function of $\lambda$ and whose 
argument is $k$ \cite{Cloizeaux66} (see the Appendix). 
Using a different method, the dispersion $\varepsilon_{c,0}(q)$ 
was previously derived  by Des Cloizeaux and Gaudin \cite {Cloizeaux66}.

The ground state is characterized by both $N_{c,\gamma >0}=0$ and
a compact symmetric distribution of the numbers $I_{c,0}$ 
around zero \cite{Yang66}.
We represent this distribution by
$N_{c,0}^0(q)$. According to Table \ref{Tba}, 
low-energy excitations can occur in the $c,0$ band 
for values of $n_f<1/2$. These excitations correspond to distributions of
the numbers $I_{c,0}$
different from the $N_{c,0}^0(q)$ distribution. 
We represent these distributions
by $N_{c,0}(q)=N_{c,0}^0(q)+\delta N_{c,0}(q)$,
where $\delta N_{c,0}(q)$ describes the deviations
of the distribution $N_{c,0}(q)$ relative to $N_{c,0}^0(q)$. From
this simple picture, we expect the system to behave
in the same way as a free fermionic system for low temperature. 

There is, however, forward scattering among the
pseudoparticles, which is not present in a free fermionic system. This interaction is described by the $f-$function
$f_{c,\gamma;c,\gamma'}$, which is given by

\end{multicols}
\vspace{-0.6truecm}
\noindent\makebox[8.8truecm]{\hrulefill}
\widetext
\begin{eqnarray}
\frac 1{2\pi}
f_{c,\gamma;c,\gamma\,'}(q,q\,') =  v_{c,\gamma}(q)
\Phi_{c,\gamma;c,\gamma\,'}(q,q\,')+v_{c,\gamma\,'}(q\,')
\Phi_{c,\gamma\,';c,\gamma}(q\,',q) \nonumber \\
 +  \sum_{j=\pm 1} 
\sum_{\gamma\,''}
v_{c,\gamma\,''}
\Phi_{c,\gamma\,'';c,\gamma}(jq_{Fc
,\gamma\,''},q)\Phi_{c,\gamma\,'';c,\gamma'}
(jq_{Fc,\gamma\,''},q\,') \, ,
\label{ffun}
\end{eqnarray}
\vspace{-0.4truecm}
\hspace*{\fill}\makebox[8.8truecm]{\hrulefill}
\begin{multicols}{2}
\narrowtext\noindent
In
equation\, (\ref{ffun})
$\Phi_{c,0;c,0}(q,q\,')$ are pseudoparticle 
phase-shifts
due to forward scattering pseudoparticle-pseudoparticle collisions \cite{Nuno97b}. In the Appendix, we give the integral equations obeyed by
the phase shifts $\Phi_{c,\gamma;c,\gamma\,'}(q,q\,')$.

These $f-$function control the behavior 
of the correlation function exponents, as is the
case of the conductivity exponents 
\cite{Carmelo98p}. The $f-$functions
also control the energy eigenvalues $E_n$, which are written
as functionals of $\delta N_{c,0}(q)$ and are of the form
\begin{equation}
E_n-E_0=E^{(1)}+E^{(2)}+({\rm higher \hspace{.1cm}order \hspace{.1cm}terms})\,,
\label{e1e2}
\end{equation}
with
\begin{equation}
E^{(1)} = \sum_{q,\gamma}
\varepsilon_{c,\gamma}(q)\delta N_{c,\gamma}(q) \, ,
\label{h1}
\end{equation}
and 
\begin{equation}
E^{(2)}={1\over {N_a}}\sum_{q,\gamma} 
\sum_{q\,',\gamma'}{1\over 2}f_{c,\gamma;
c,\gamma\,'}(q,q') \delta N_{c,\gamma}(q)
\delta N_{c,\gamma\,'}(q') \, ,
\label{h2}
\end{equation}
and $E_0$ is the ground state energy.
Similar results to Eqs. (\ref{e1e2}), (\ref{h1}), and (\ref{h2}) have been derived for the Hubbard model
\cite{Nuno97b}, and as long as $\delta_{c,\gamma}/N_a$
--the number of excited pseudopartocles-- is small the
higher order terms are negligible.

The relation between the gapless $c,0$ excitations and the 
original fermionic problem is best understood in the Luttinger liquid paradigm
\cite{Haldane81,Haldane91}. These excitations correspond 
to the density waves obtained by bosonization of 
Hamiltonian (\ref{spinless}), which, in turn, are 
generated by the low energy particle-hole processes 
around the non-interacting Fermi surface.

A different situation occurs if $n_f=1/2$. For this canonical ensemble,
Table \ref{Tba} shows that the number of available $I_{c,0}$ numbers is equal
to the number of $c,0$ pseudoparticles occupying those orbitals. As
a consequence, no low-energy excitations can occur
in the $c,0$ band. The lowest energy excitation requires that one
$N_{c,1}$ pseudoparticle is created, which costs a finite energy. In this
process two available states appear in the $c,0$ band and
therefore the remaining particles 
can now undergo {\it particle-hole}
excitations within the $c,0$ band.

\begin{figure}[htbp]
\epsfxsize=8.0 truecm 
\epsfysize=8.0 truecm
\centerline{\epsffile{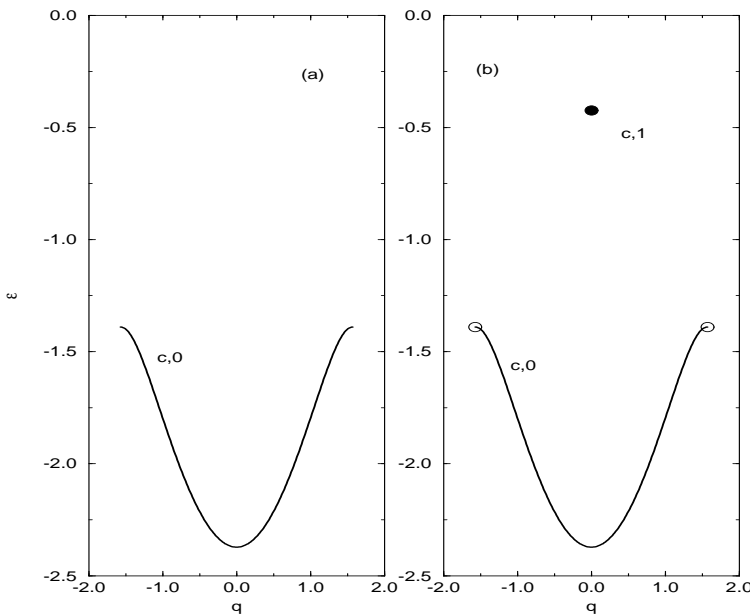}}
\vspace{1cm}
\caption{Plot of the filled $c,0$ 
band (corresponding to half-filling of the spinless fermions) 
and lowest energy excitation to the $c,1$ 
band (creating a pseudoparticle in the $c,1$ 
band and leaving two pseudoholes in the $c,0$ 
band in accordance with Eq. (\ref{sumrule})).}
\label{bandfig}
\end{figure}

For $V\gg J$ the $c,1$ excitation can be seen
as an excitation to an ``upper Hubbard band" and is related, in the original
lattice, to the creation of a pair of two nearest-neighbor occupied
sites. In this
limit, the gap $\Delta_1$ has the asymptotic form $\Delta_1\simeq V-2J$.
This excitation process is the most important contribution for the
zero-temperature behavior of $\sigma_{reg}(\omega,0)$ 
at half-filling \cite{Carmelo98p}.
In Fig. \ref{bandfig} we plot the $c,0$ band at 
half-filling and represent the lowest energy
excitation to the $c,1$ band (represented by a point).

From
this description, we see
that there is a mapping between an insulator due to correlations in the
fermionic picture and a {\it band} insulator in the 
pseudoparticle representation.
This mapping holds, in this model, for all $V>J$ and is not restricted to
the ``upper Hubbard band" strong-coupling argument given above. The same
mapping holds for the Hubbard model \cite{Nuno97b,Imada95,Shankar}. 

\subsection{$D_{\lowercase{n}}$ for $n_f<1/2$ and $T\ll
v_{c,0}q_{c,0}$}
\label{thermoapp}

Due to the presence of the gaps $\Delta_{\gamma}$ 
the behavior
of the curvature of levels $D_{\lowercase{n}}$ at low
temperatures and away from half filling is
mainly controled by the $c,0$ excitations. Therefore in
the calculation of $D_{\lowercase{n}}$ below
we ignore the contributions from the states with
$c,\gamma >1$ excitations (these will play a major role
only at half filling).
 
We now make these considerations more quantitative. For the calculation
of Eq.\,(\ref{dt}), we need to compute the curvature of levels,
$D_n$. The eigenstates characterized
by the $q=2\pi I_{c,0}/N_a$ numbers and the corresponding eigenenergies
$E_n$ are expressed as functionals of the distributions
$N_{c,0}(q)$ of these quantum numbers. We write
$N_{c,0}(q)=N_{c,0}^0(q)+\delta N_{c,0}(q)$,
where $N_{c,0}^0(q)$ is a compact distribution of all
$I_{c,0}$ numbers symmetrically distributed around zero. If the deviation
$\delta N_{c,0}(q)$ is small, we can expand the relevant quantities
in $\delta N_{c,0}(q)$. 
Using the BA equations with twisted boundary conditions \cite{Shastry90},
we next derive simple 
equations for $j_n$ and $D_n$ as functions of $N_{c,0}(q)$.

Following references\,\cite{Stafford90,Castella95,Nuno97a,Nuno97b,Nuno97c}, the
general expression for $j_n$ reads
\begin{equation}
        j_n=J\sinh^2\lambda
        \sum_{q}N_{c,0}(q)
        \frac{\sin[\Lambda_{c,0}(q)]\Lambda_{c,0}^{\phi}(q)}
        {\{\cosh\lambda-\cos[\Lambda_{c,0}(q)]\}^2}\,,
\label{jex}
\end{equation}
where $\Lambda_{c,0}(q)$ is defined by taking the
limit $N_a\rightarrow \infty$ in Eq.\,(\ref{bethe})
and $\Lambda_{c,0}^{\phi}(q)$ is the first 
derivative of $\Lambda_{c,0}^j$ with respect to $\phi$
such that
\begin{equation}
\Lambda^{\phi}_{c,0}(q)=
\lim_{N_a\rightarrow\infty} \left .\frac{d\,\Lambda^j_{c,0}
(2\pi I^j_{c,0}/N_a)}{d(\phi/N_a)}
\right\vert_{\phi=0}\,.
\label{deriv}
\end{equation}
We stress that Eq. (\ref{jex}) is the exact mean value of the
current operator (\ref{cur}) in any 
eigenstate of the Hamiltonian (\ref{spinless}) of the
low energy sector (only $c,0$ excitations included).
That is, Eq. (\ref{jex}) holds for any $N_{c,0}(q)$.
By methods equivalent to those used previously
for the Hubbard model \cite{Nuno97a,Nuno97c} and for
small densities of excited $c,0$ pseudoparticles the FDE gives the following result
for the mean value of the current operator\ (\ref{jex}):
\begin{equation}
        j_n=\sum_q 
        \delta N_{c,0}(q)j_{c,0}(q)\,,
\label{exactj}
\end{equation}
with the spectrum $j_{c,0}(q)$ given by
\begin{eqnarray}
        j_{c,0}(q)&=&
        v_{c,0}(q)Q_{c,0}^{\Phi,0}(q)+
      \sum_{j=\pm 1}
        v_{c,0}
        \Phi_{c,0;c,0}(j,q)Q_{c,0}^{\Phi,0}(j)\,,
\nonumber\\
        Q_{c,0}^{\Phi,0}(q)&=&
       1+\sum_{j=\pm 1}
      j\Phi_{c,0;c,0}(q,jq_{F,0})\,,
\label{func}
\end{eqnarray}
$v_{c,0}(q)=d\varepsilon_{\alpha,0}(q)/dq$,
and $v_{c,0}=v_{c,0}(q_{Fc,0})$. 

Equation (\ref{func}) has a simple physical meaning, since it is the sum
of the velocity plus a dragging term due to the collisions among the
pseudoparticles (similar results have been derived for the Hubbard 
model \cite{Carmelo92,Nuno97c}). This picture is very similar to
that occurring in a Fermi liquid \cite{Baym}.
The curvature of the levels $E_n$ can also be computed. For any
low-energy eigenstate, its curvature $D_n$ is given by
\end{multicols}
\vspace{-0.6truecm}
\noindent\makebox[8.8truecm]{\hrulefill}
\widetext
\begin{eqnarray}
        D_n&=&-J\frac{\sinh^2\lambda} {2}
        \sum_{q}N_{c,0}(q)\left (
        -\frac{\sin[\Lambda_{c,0}(q)]\Lambda_{c,0}^{\phi\phi}(q)
        +\cos[\Lambda_{c,0}(q)][\Lambda_{c,0}^{\phi}(q)]^2}
        {\{\cosh\lambda-\cos[\Lambda_{c,0}(q)]\}^2}
        +
        2\frac{\sin^2[\Lambda_{c,0}(q)]
        [\Lambda_{c,0}^{\phi}(q)]^2}
        {\{\cosh\lambda-\cos[\Lambda_{c,0}(q)]\}^3}
        \right )\,,
\label{dex}
\end{eqnarray}
\vspace{-0.4truecm}
\hspace*{\fill}\makebox[8.8truecm]{\hrulefill}
\begin{multicols}{2}
\narrowtext\noindent
where $\Lambda_{c,0}^{\phi\phi}(q)
$ is  the second derivative of $\Lambda_{c,0}^j$ with respect
to $\phi$ in the
sense of Eq. (\ref{deriv}).

For small densities of excited $c,0$ pseudoparticles 
using the FDE
allows us to express Eq.\, (\ref{dex}) as

\begin{eqnarray}
        D_n&=&D_0+ \frac 1{2}\sum_q 
        \delta N_{c,0}(q)D_{c,0}(q)\,,\nonumber\\
        D_0&=&\frac {Na} {4\pi}\sum_{j=\pm 1}v_{c,0}\left [
        Q_{c,1}^{\Phi,0}(jq_{F,0}) \right ]^2=
        \frac{N_a}{2\pi}v_{c,0}(\xi_{c,0})^2\,,
\label{curvature}
\end{eqnarray}
where $D_0/N_a$ is the zero-temperature 
charge stiffness and $\xi_{c,0}$ is the 
dressed charge. Identical relations have been 
obtained by other authors for other integrable models \cite{Kawakami91}.
It is simple to
show that $D_0=j_{c,0}(q_{F,0})/(2\pi)$, and similar results have been
derived for the Hubbard model \cite{Carmelo92,Nuno97c,comm1}. The function
$D_{c,0}(q)$ is an even function of $q$ and is a combination of
$v_{c,0}(q)$, $\Phi_{c,0;c,0}(q,q')$, and their derivatives. 
We give
its general form in the Appendix. 

Close to half filling, $D_{c,0}(q)\rightarrow d v_{c,0}(q)/dq$ is the curvature of
the energy band and $D_0\rightarrow 0$, as computed
by Haldane \cite{Haldane81b}.  This latter result is a consequence
of the vanishing velocity \cite{Haldane81b}
$v_{c,0}(q)$ at $q=\pi/2$, as can be seen directly from Eq.\,(\ref{dispsf}).
This signals the zero temperature metal-insulator transition in this model.
The spectrum of the curvature, $D_{c,0}(q)$, has a very simple form at
close to half filling, when compared to its general expression given in the Appendix. 
It bears some resemblance to the result for the independent fermion gas, 
but here 
$d v_{c,0}(q)/dq$ refers to the $c,0$ energy excitations of the
many-body system.
\section{Asymptotic BA Results for Arbitrary
System Sizes} 
\label{asymptotics}

In the previous section we have considered the curvature of levels in the thermodynamic limit and we used
the FDE formalism to compute $j_n$ and $D_n$ for the low energy states. We now consider the $N_a$ 
dependence of $j_n$ and $D_n$. The calculation
of $j_n$ and $D_n$ for all eigenstates for finite-size
$N_a$ is a difficult problem in general, but in the limits $V=\infty$ and $V\gg t$, it can be performed explicitly.
Our starting point is the $V=\infty$ solution which we will use to compute the asymptotic corrections of order
$1/V$.

\subsection{$V=\infty$ solution}

The limit $V\rightarrow\infty$ is a well defined limit
of the BA equations and therefore is suitable for classical
perturbation theory. In the limit $V=\infty$ the 
BA equations (\ref{bethe}) and (\ref{eig}) can be written as
\begin{eqnarray}
        &&N_a\Lambda_{c,\gamma}^{j,\infty}=
        2\pi I_{c,\gamma}^j
+(\gamma+1)\phi\nonumber\\
&&+\sum_{\gamma'=0}\sum_{p=1}
        [\gamma\gamma'p]
         \sum_{j'=1}^{N_{c,\gamma\,'}}
        (\Lambda_{c,\gamma}^{j,\infty}-
        \Lambda_{c,\gamma'}^{{j',\infty}})\,,
\hspace{.4cm}j=1,\ldots,N_{c,\gamma}
\label{betheinfty}
\end{eqnarray}
and
\begin{equation}
E_n=-J\sum_{j=1}^{N_{c,0}}\cos \Lambda_{c,0}^{j,\infty}\,,
\label{einf}
\end{equation}
respectively. In this limit we see that the set of Eqs. (\ref{bethe}) transform into a set of coupled linear algebraic
systems.
This result shows that if $V=\infty$ only
the $\Lambda_{c,0}^{j,\infty}$ parameters are of importance, in what
concerns the energy eigenvalues, but, nevertheless, Eq. (\ref{betheinfty}) is meaningfull for all $\Lambda_{c,\gamma}^{j,\infty}$. Considering
only the $c,0$ excitations, Eq. (\ref{betheinfty}) can be solved exactly and reads
\begin{eqnarray}
\Lambda_{c,0}^{j,\infty}=\frac{\phi}{N_a}+\frac{2\pi}{N_a-N_f}I_{c,0}^j
+\nonumber\\
\left(\frac{2\pi}{N_aN_f}-\frac{2\pi}{N_f(N_a-N_f)}\right)\,.
\sum_{i=1}^{N_f}I^i_{c,0}
\label{inftysol}
\end{eqnarray}
The
picture that emerges from this solution is essentialy that of non-interacting particles, in an effective ring of size $N_a-N_f$ \cite{Kusmartsev92}. For this case,
the curvature of levels is simply given by
\begin{equation}
D_n=\frac{J}{2N_a}\sum_{\{I_{c,0}^j\}_n}\cos
\Lambda_{c,0}^{j,\infty}\,
\label{cruvinf}
\end{equation}
and  
the zero temperature charge stiffness can be computed using Eqs. (\ref{inftysol})
in Eq. (\ref{cruvinf}) giving \cite{Kusmartsev92,comm}
\begin{equation}
D(0)=\frac{J}{2N_a}\frac{\sin(\pi N_f/(N_a-N_f))}{\sin(\pi /(N_a-N_f))}\,.
\end{equation}
 Therefore, as long as the system is not
at half filling we expect ideal conductivity behavior at
finite temperature, in agreement with the picture we developed
in the previous section for the thermodynamic limit.

The set of equations (\ref{betheinfty}) can be solved for
arbitrary occupancy of the numbers $N_{c,\gamma}$ and this
solution can be used as a starting point to an asymptotic
calculation of the energy levels $E_n$ and
its corresponding curvature of levels
$E_n$ as function of the system size $N_a$. The curvature of levels $D_n$ depends
in a crucial way on the flux dependence of the parameters
$\Lambda_{c,\gamma}^j$. We now determine this dependence by
solving  Eq. (\ref{betheinfty}).

From Eq. (\ref{sumrule}) the highest possible 
occupied band is that
characterized by $\gamma_f=N_f-1$. Let us consider an arbitrary occupancy of the numbers $N_{c,\gamma}$ compatible
with Eq. (\ref{sumrule}). This configuration is always of the
form
\begin{equation}
N_{c,0}=\nu_0; N_{c,1}=\nu_1, \ldots,
;N_{c,\bar m}=\nu_{\bar m} 
\end{equation}
and
\begin{equation}
N_{c,1+\bar m}=0; \ldots;
N_{c,\gamma_f}=0\,,
\end{equation}
with the possibility that $\bar m=\gamma_f$. In order to solve our algebraic coupled linear systems we first need to obtain
$\Lambda_{c,\bar m}^{j,\infty}$. Incidentally
we only need to compute the sum
\begin{equation}
\sum_{j=1}^{\nu_{\bar m}}
\Lambda_{c,\bar m}^{j,\infty}\,,
\label{sumbar}
\end{equation}
which, as can be seen from Eq. (\ref{betheinfty}), 
is present in all the equations that define $\Lambda_{c,\gamma}^{j,\infty}$, with $\gamma < \bar m$.  Using the equation that defines
$\Lambda_{c,\bar m}^{j,\infty}$, we obtain

\begin{eqnarray}
(N_a&-&2N_f+2(\bar m+1) \nu_{\bar m})\sum_{j=1}^{\nu_{\bar m}}\Lambda_{c,\bar m}^{j,\infty}=
2\pi\sum_{j=1}^{\nu_{\bar m}}I_{c,\bar m}^j 
\nonumber\\
&+&\nu_{\bar m}(1+\bar m)\phi-
\nu_{\bar m}\sum_{\gamma=0}^{\bar m -1}\sum_{j=1}
^{\nu_{\gamma}}
2(\gamma +1)\Lambda_{c,\gamma}^{j,\infty}\,.
\end{eqnarray}
Substituting the above sum in Eq. (\ref{betheinfty}) for a given $\gamma<\bar m$ we obtain
\end{multicols}
\vspace{-0.6truecm}
\noindent\makebox[8.8truecm]{\hrulefill}
\widetext
\begin{eqnarray}
        (N_a-2(\gamma+1)\nu_{\bar m})\Lambda_{c,\gamma}^{j,\infty}=
        2\pi I_{c,\gamma}^j
+\frac{(N_a-2N_f)(\gamma+1)\phi}{N_a-2N_f+2(\bar m+1)}
+\sum_{\gamma'=0}^{\bar m -1}\sum_{p=1}
        [\gamma\gamma'p]
         \sum_{j'=1}^{\nu_{\gamma\,'}}
        (\Lambda_{c,\gamma}^{j,\infty}-
        \Lambda_{c,\gamma'}^{j',\infty})\nonumber\\
		-\frac{2(\gamma+1)}{N_a-2N_f+2(\bar m+1)}\left(
2\pi\sum_{j=1}^{\nu_{\bar m}}I_{c,\bar m}^j -
\nu_{\bar m}\sum_{\gamma'=0}^{\bar m -1}\sum_{j=1}
^{\nu_{\gamma\,'}}
2(\gamma +1)\Lambda_{c,\gamma'}^{j,\infty}\right )
\hspace{1cm} (\gamma <\bar m;\bar m >0)\,.
\label{bethesol}
\end{eqnarray}
\vspace{-0.4truecm}
\hspace*{\fill}\makebox[8.8truecm]{\hrulefill}
\begin{multicols}{2}
\narrowtext\noindent
It is simple to prove that if only the $c,0$ excitations
have non-zero occupancy then Eq. (\ref{bethesol}) gives
Eq. (\ref{inftysol}). Moreover we see that the flux term, due
to the presence of $c,\gamma>0$ excitations is modified
as follows
\begin{equation}
	\phi\rightarrow\Phi^{eff}=\frac{(N_a-2N_f)\phi}{N_a-2N_f+2(\bar m+1)}
\label{phieff}
\end{equation}
The above equation shows, in the regime $V\gg t$,
that if $c,\gamma>0$ excitations are allowed, the 
flux felt by the several $c,\gamma$ excitations is
$\Phi^{eff}$, with the exception of the $c,\bar m$
excitation. Even more important is
the fact that the effective flux $\Phi^{eff}$ is zero at half
filling. This property turns out to be important in
Section \ref{vggt} where we consider the asymptotic solution
$V\gg t$.
\subsection{$V\gg t$ solution: $J^2/V$ corrections}
\label{vggt}

Although the set of equations (\ref{betheinfty}) has
a solution for $V=\infty$, the energy of the eigenstates
with non-zero occupancy of the $c,\gamma >0$ bands is
infinite. In the regime $V\gg t$ the energy
of the eigenstates with finite occupancy of the $c,\gamma >0$ bands is given by
\begin{equation}
E_n=-J\sum_{j=1}^{N_{c,0}}\cos \Lambda_{c,0}^{j,\infty}+J\cosh \lambda \sum_{\gamma=1}\gamma N_{c,\gamma}
\label{energyasymp}
\end{equation}
with $\Lambda_{c,0}^j$ given by Eq. (\ref{bethesol}).
Therefore, up the order where $J^2/V$ corrections are neglected \cite{comm2}, we have two different situations:
(a) away from half filling, the $c,0$ holes that
appear in the $c,0$ band due to the creation of $c,\gamma>0$
excitations fell the effective flux $\Phi^{eff}$ but they
still behave as {\it independent particles} and therefore
the system display ideal (infinite) conductivity at finite
temperature (this analysis agrees with the thermodynamic
limit solution consider in Section \ref{thermoapp}); (b) at
half filling $\Phi^{eff}=0$ and because, up to this order, $\Lambda_{c,0}^j$ is independent of 
$\Lambda_{c,\bar m}^j$ all the derivatives of 
Eq. (\ref{energyasymp}) in order to
the flux $\phi$ are identically zero and the system
behaves as ideal insulator (or at least does not display
infinite conductivity see Eq. (\ref{cond})).

We now compute the $1/V$ corrections
to Eqs. (\ref{bethe}) and (\ref{eig}). To keep the
calculations as simple as possible we adopt a low
temperature scheme by considering a single type 
of gapped $c,\gamma$ excitations, namely the $c,1$ 
excitation \cite{Takahashi97}, which correspond
to the gap $\Delta_1=V-2J$ (negleting
$1/V$ corrections). We show below that this
approach leads again to an ideal insulator behavior
at half filling.
We stress that our approach is equivalent to that of Fujimoto and Kawakami \cite{Kawakami97}, for
the Hubbard model,
using the thermodynamic BA. 

Our starting point is the result for $\Lambda_{c,0}^j$
and $\Lambda_{c,1}^j$  for $V=\infty$ which we have
denoted
by $\Lambda_{c,0,1}^{j,\infty}$. In the limit $V \gg t$
we look for a solution to the parameters $\Lambda_{c,0,1}^j$ of the form $\Lambda_{c,0,1}^j=\Lambda_{c,0,1}^{j,\infty}+
\Lambda_{c,0,1}^{j,1/V}$ (with $V=J\cosh \lambda$),
where $\Lambda_{c,0,1}^{j,1/V}$ is the $1/V$ correction
to $\Lambda_{c,0,1}^j$.
By straightforward expansion of Eq. (\ref{eig}) we obtain 
\begin{eqnarray}
E_n=&-&J\sum_{j=1}^{N_{c,0}}\left (\cos \Lambda_{c,0}^{j,\infty}+\frac{J\cos^2 \Lambda_{c,0}^{j,\infty}}
{V}
-\Lambda_{c,0}^{j,1/V}\sin \Lambda_{c,0}^{j,\infty}
\right)+\nonumber\\
&-&J^2\sum_{j=1}^{N_{c,1}}\frac{\cos \Lambda_{c,1}^{j,\infty}}
{2V}+
J^2(V-V^{-1})N_{c,1}\,.
\label{easympvt}
\end{eqnarray}
The above equation tell us that we need to obtain
$\Lambda_{c,0}^{j,1/V}$ to compute $E_n$ consistently up to order $1/V$. From
Eq. (\ref{bethe}) we obtain 
\begin{equation}
\Lambda_{c,0}^{i,1/V}=-\frac{J}{V}\sin{\Lambda_{c,0}^{i,\infty}}-
\frac{J}{N_aV}\sum_{j=1}^{\nu_2}\sin(\Lambda_{c,1}^{j,\infty}-\Lambda_{c,0}^{i,\infty})\,.
\label{lambvt}
\end{equation}
From the study of the $\Lambda_{c,\gamma}^{j,\infty}$
solution of the previous Section it follows that Eqs. (\ref{easympvt}) and (\ref{lambvt}) only depend
on the flux $\phi$ through the parameter
$\Lambda_{c,1}^{j,\infty}$, which in the general case corresponds to $\Lambda_{c,\bar m}^{j,\infty}$. 

We now prove that the terms $\sum_j^{\nu_1}\sin(\Lambda_{c,1}^{j,\infty}-\Lambda_{c,0}^{i,\infty})$ and $\sum_j^{\nu_1}
\cos(\Lambda_{c,1}^{j,\infty})$ are indeed flux independent
at half filling, although $\Lambda_{c,1}^{j,\infty}$ is not. This would mean that any thermodynamic
BA calculation of the type developed for the Hubbard
model \cite{Kawakami97} would give a zero finite-temperature
charge stifness. We start by obtaining $\Lambda_{c,1}^{j,\infty}$ at half filling. For the
case we are considering this is simple to obtain
(we give the general solution for $\Lambda_{c,\bar m}^{j,\infty}$ in the Appendix \ref{apb}) and reads
\begin{equation}
\Lambda_{c,1}^{i,\infty}=\frac{2\pi I_{c,1}^i}{\nu_1}
-\frac 1 {2\nu_1}\sum_{j=1}^{\nu_0}\Lambda_{c,0}^{j,\infty}
+\frac{\phi}{2\nu_1}-\frac{3\pi}{2\nu_1^2}\sum_{j=1}^{\nu_1}
I_{c,1}^j\,.
\label{barsol}
\end{equation}
Using the above result together with Eq. (\ref{interval}),
it is simple to see that both $\sum_j^{\nu_2}\sin(\Lambda_{c,1}^{j,\infty}-\Lambda_{c,0}^{i,\infty})$ and $\sum_j^{\nu_2}
\cos(\Lambda_{c,1}^{j,\infty})$ are zero at half filling.

Hence we see that the curvature of levels $D_n$
is  identically zero and therefore the charge stiffness
$D(T)$ is also zero. We remark here that
our result does not dependent on fact that
only $c,0$ and $c,1$ excitations have been considered. The main conclusion
is that for arbitrary values of the numbers
$N_{c,\gamma}$, consistent with the Eq. (\ref{sumrule}), the
highest excitation band $c,\bar m$ is always completely 
filled (see Table \ref{Tba} ) and, in addition,
is the only band that is flux dependent (up to order
$1/V$). As result we always have $D_n=0$ and a zero value
for $D(T)$ at half filling.

\section{Discussion and Conclusions}
\label{lconc}

We have proved, up to order $1/V$, that the charge
stiffness of the $t-V$ model is zero at any temperature
for the half-filled band case. This means that the system
does not present balistic transport of charge in this regime.
Our result proves a conjecture put forward first by Zotos and
Prelov\v{s}ek \cite{Zotos96} based on exact 
diagonalization of small rings. If we go to the next order
$1/V^2$ there will be contributions to the curvature
of levels $D_n$ that vanishes in the thermodynamic limit
\cite{comment}. Furthermore, our results show that
the Mazur inequality --$D(T)\ge 0$-- derived in Ref. \onlinecite{Naef97}, holds as an equality.

Away from half filling, $D(T)$ is mainly 
controlled by $D_0$ for low $T$ (that is, $T\ll v_{c,0}q_{Fc,0}$).
Since the states with finite $c,\gamma >0$ pseudoparticle
occupancy
occupied have energy gaps relative to the ground state, 
only the $c,0$ pseudoparticles contribute to the charge transport at finite but small
temperatures. This implies that for low $T$ the model can be thought as 
a {\it gas} of $c,0$ pseudoparticles, and therefore $D(T)$ is finite and the
system is an {\it ideal conductor} in
agreement with Ref. \cite{Zotos96}.
Of course, the curvature of levels $D_n$ is not 
as simple as in a true fermionic gas, due to the forward scattering
among the pseudoparticles (these scattering processes are controlled by the
phase shifts as in the Hubbard model \cite{Nuno97b}).
Nevertheless, the forward scattering among the pseudoparticles will
only renormalize the $b$ coefficient in $D(T)=D_0-bT^2$ away from its 
non-interacting value. This picture is supported by 
the analysis of the BA equations  (\ref{bethe})
in the extreme correlated case 
of $V=\infty$, which, for large $N_a$, can be consider as {\it free} fermions in 
an effective lattice of size $L_{eff.}=N_a-N_f$.

Our results also provide insight into spin transport in the spin-$1/2$
Heisenberg chain, and similar conclusions to those above hold for the
temperature dependence of the spin stiffness. The regime $V > J$ corresponds
to the {\it Ising symmetry} of the spin-$1/2$ anisotropic 
Heisenberg chain. In this sector, the exchange coupling in the $z$ direction 
is 
larger than the coupling in the $xy$ plane. The half-filled band (non-half 
filled band) case  in the spinless fermions model  
corresponds to the zero 
magnetic field
(finite magnetization) case in the Heisenberg chain. 

As a final comment, we believe that the anomalous transport properties
exhibited by some of the models solvable by the BA arise
from the existence of only forward scattering among the pseudoparticles
at all energy scales. As discussed in Sec. \ref{introdd}, this is a consequence
of the integrability in these models and is not expected to occur in
non-integrable many-body systems.

\section*{ACKNOWLEDGMENTS}
We are grateful to Ant\'{o}nio Castro Neto and
Felix Naef
for reading the earlier version of this
manuscript and for their criticism. 
N.M.R. Peres wants to thank the hospitality of
the University of Illinois where this work was begun, the
Gulbenkian and Luso-American
Foundations for financial support, and Lu\'{\i}s Miguel Martelo
and Miguel Ara\'{u}jo
for useful discussions. The support of the US National
Science Foundation under grant NSF DMR 97-12765 is gratefully acknowledged. 
\appendix
\section{General Equations for the Phase shifts, Energy Bands, and
$D_{\lowercase{c},0}(\lowercase{q})$}

In this appendix we give the general equations for the
phase shifts $\Phi_{c,\gamma;c,\gamma'}(q,q')$ and for the bands 
$\varepsilon_{\alpha,\gamma}(q)$. Following standard methods (see,
e.g., Ref. \cite{Nuno97b}) we obtain 
\begin{eqnarray}
&&\bar{\Phi}_{c,\gamma;c,\gamma'}(\Lambda_{c,\gamma},
\Lambda_{c,\gamma'})=\nonumber\\
&&\sum_{p=1}\left\{
\frac{[\gamma\gamma' p]}{\pi}\arctan\left( A_p\tan\frac{\Lambda_{c,\gamma}-
\Lambda_{c,\gamma'}}{2}\right)-\right .\nonumber\\
&-&\left . \int_{-x_0}^{x_0}dx \frac{[\gamma 0 p]}{\pi}
\frac{A_p\bar{\Phi}_{c,0;c,\gamma'}(x,\Lambda_{c,\gamma'})}{1+A^2_p-(A^2_p-1)\cos(x-
\Lambda_{c,\gamma})}
\right\}
\end{eqnarray}
and
\begin{eqnarray}
&&\varepsilon^0_{c,\gamma}(q)=-J\frac{\sinh[(\gamma+1)\lambda]\sinh\lambda}
{\cosh[(\gamma+1)\lambda]-\cos[\Lambda_{c,\gamma}^0(q)]}
+\nonumber\\&+&J\int_{-x_0}^{x_0}dx 
\frac{\sinh^2\lambda\sin(x)\bar{\Phi}_{c,0;c,\gamma}[x,\Lambda^0_{c,\gamma}(q)]}{\{\cosh[(\gamma
+1)\lambda]-\cos(x)\}^2}\,,
\end{eqnarray}
where 
$\Phi_{c,\gamma;c,\gamma'}(q,q')=\bar{\Phi}_{c,\gamma;c,\gamma'}[\Lambda^0_{c,\gamma}(q),\Lambda^0_{c,\gamma'}(q\,')]$, $x_0=\Lambda^0_{c,0}(q_{Fc,0})$ and $\Lambda^0_{c,0}(q)$ is 
$\Lambda_{c,0}(q)$ in the ground state and it can be written in terms of the phase shifts 
$\Phi_{c,\gamma;c,\gamma'}(q,q')$ as
\begin{eqnarray}
&&2\arctan\left( A_{\gamma}\tan\frac{\Lambda^0_{c,\gamma}(q)}{2}\right)=
q-\nonumber\\
&-&2A_0\int_{-x_0}^{x_0}dx
\frac{\bar{\Phi}_{c,0;c,\gamma'}[x,\Lambda^0_{c,\gamma'}(q)]}{1+A^2_0-(A^2_0-1)\cos x}\,.
\end{eqnarray}
These three equations can be solved in closed form. In general,
only a numerical
solution is possible. 
For $n_f=1/2$ one has $q_{Fc,0}=\pi/2$ and $\Lambda^0_{c,0}(q_{Fc,0})=\pi$,
and this set of integral equations can be solved by Fourier series. The results
for the phase shifts and for the bands are

\begin{eqnarray}
&&\bar{\Phi}_{c,0;c,\gamma'}[\Lambda^0_{c,0}(q),\Lambda^0_{c,\gamma'}(q\,')]=\nonumber\\
&&\sum_{n=1}^{\infty}\sum_{p=1}
\frac{[0\gamma' p]}{n\pi}\sin[n(\Lambda^0_{c,0}(q)-\Lambda^0_{c,\gamma'}(q\,'))]\frac{e^{-
pn\lambda}}{1+e^{-2n\lambda}}
\end{eqnarray}
and
\begin{eqnarray}
\varepsilon_{c,0}(q)&=&-J\sinh\lambda-
J\sum_{n=1}^{\infty}\frac{\cos[n\Lambda^0_{c,0}(q)]}
{\cosh(n\lambda)}\nonumber\\
\varepsilon_{c,\gamma}(q)&=&-J\sinh\lambda\,,
\end{eqnarray}
respectively. The relation of these expressions with Eq.(\ref{dispsf}) are 
obtained by using the {\it delta} and {\it amplitude} functions \cite{Table}. 
The parameter $k$ is a function of $\lambda$ \cite{Cloizeaux66}
$k=4 \left[ \sum_{n=1} \exp \left( -\lambda (n-0.5)^2 \right) \right]^2/
\left[1+2 \sum_{n=1}
\exp \left( -\lambda n^2 \right) \right]^2$.
Both the phase shifts and the velocities enter in the general expression
for $D_{c,0}(q)$ 
(see Eq.\,(\ref{curvature})). The general form of $D_{c,0}(q)$ reads

\begin{eqnarray}
        &&D_{c,0}(q)=\frac d{dq}\left[
        v_{c,0}(q)(Q^{\phi,0}_{c,0}(q))^2     \right]+
        v_{c,0}(q)W^{\phi\phi,0}(q)\nonumber\\
&+&
        \sum_{j=\pm 1} \left \{
        2v_{c,0}(q_{Fc,0})
        \Phi_{c,0;c,0}^{\phi}
        (jq_{Fc,0},q)     Q^{\phi,0}_{c,0}(jq_{Fc,0})
        \right .\nonumber\\
&+&
         j\left . \frac d{dq\,'}
        \left[v_{c,0}(q\,')\Phi_{c,0;c,0}
        (q,q\,') [Q^{\phi,0}_{c,0}(q\,')]^2
        \right]\right\vert_{q\,'=jq_{Fc,0}}
        \nonumber\\
&+&
        \left .
        v_{c,0}(q_{Fc,0})
        \Phi_{c,0;c,0}
        (jq_{Fc,0},q)     W^{\phi\phi,0}_{c,0}(jq_{Fc,0})\
\right \}\,.
\label{fd}
\end{eqnarray}
The functions $W^{\phi\phi,0}_{c,\gamma}(q)$ and
$\Phi_{c,\gamma;c,\gamma'}^{\phi}(q,q\,')$ are 
given by
\begin{eqnarray}
        W^{\phi\phi,0}_{c,0}(q)=\sum_{0}\sum_{j\pm 1}j
[Q^{\phi,0}_{c,0}(jq_{Fc,0})]^2\nonumber\\
\hspace{3cm}
\left . \frac {d \Phi_{c,0;c,0}(q,q\,')}{dq\,'}\right\vert_{q\,'=jq_{Fc,0}}\,,
\label{fw}
\end{eqnarray}
and
\begin{eqnarray}
&&\Phi_{c,0;c,0}^{\phi}(q,q\,')= 
\frac {d \Phi_{c,0;c,0}(q,q\,')}{dq\,'}
Q^{\phi,0}_{c,0}(q\,')\nonumber\\
&+&
\sum_{j=\pm 1}
jQ^{\phi,0}_{c,0}(jq_{Fc,0})
\Phi_{c,0;c,0}(jq_{Fc,0},q\,')
\nonumber\\
&&\hspace{3cm}\left .\frac {d \Phi_{c,0;c,0}(q,q\,')}{dq\,'}
\right\vert_{q\,'=jq_{Fc,0}}\,,
\label{ff}
\end{eqnarray}
respectively.
\section{General solution of 
$\Lambda^{\lowercase{j},\infty}_{\lowercase{c},\bar 
{\lowercase{m}}}$ for all fillings}
\label{apb}

The equation that defines $\Lambda_{c,\bar m}^{j,\infty}$
is 
\begin{eqnarray}
        &&N_a\Lambda_{c,\bar m}^{j,\infty}=
        2\pi I_{c,\gamma}^j
+(\bar m+1)\phi\nonumber\\
&&+\sum_{\gamma'=0}^{\bar m}\sum_{p=1}
        [\bar m\gamma'p]
         \sum_{j'=1}^{N_{c,\gamma\,'}}
        (\Lambda_{c,\bar m}^{j,\infty}-
        \Lambda_{c,\gamma'}^{{j',\infty}})\,,
\label{apbetheinfty}
\end{eqnarray}
and can be cast in the form
\begin{eqnarray}
\Lambda_{c,\bar m}^{1,\infty}+\Lambda_{c,\bar m}^{2,\infty}+
\ldots+
\frac{a}{b}\Lambda_{c,\bar m}^{j,\infty}+
\nonumber\\\ldots+
\Lambda_{c,\bar m}^{\nu_{\bar m},\infty}=\frac{2\pi I^{j,\infty}_{c,\bar m}}{b}+\frac{B(\phi)}{b}
\end{eqnarray}
where $a$, $b$, and $B(\phi)$ are given by
\begin{equation}
a=N_a-2N_f+\nu_{\bar m}+2\bar m +1\,,
\end{equation}
\begin{equation}
b=2\bar m +1\,,
\end{equation}
and
\begin{equation}
B(\phi)=(\bar m +1)\phi-\sum_{\gamma=0}^{\bar m -1}
\sum_{j'=1}^{\nu_{\gamma}}2(\gamma +1 )\Lambda_{c,\gamma}
^{j',\infty}\,.
\end{equation}
This linear system of order $\nu_{\bar m}$ can be solve,
and the solution $\Lambda_{c,\bar m}^{j,\infty}$ is given by

\begin{eqnarray}
\Lambda_{c,\bar m}^{j,\infty}[&&a/b(a/b+\nu_{\bar m}-2)-\nu_{\bar m}+1]=-\frac{2\pi}{b}\sum_{j'=1}^{\nu_{\bar m}}
I^{j'}_{c,\bar m}+\nonumber\\
&&\frac{2\pi I^{j}_{c,\bar m}}{b}(a/b+\nu_{\bar m}-1)
-\frac{B(\phi)}{b}(a/b-1)\,.
\end{eqnarray}
At half filling we have $a=\nu_{\bar m}+2\bar m +1$
and $\Lambda_{c,\bar m}^{j,\infty}$ is equal to
\begin{equation}
\Lambda_{c,\bar m}^{j,\infty}=\frac {2\pi I^{j}_{c,\bar m}}
{\nu_{\bar m}}-\frac {2\bar m +1}{\bar m +1}\frac{\pi}
{(\nu_{\bar m})^2}\sum_{j'=1}^{\nu_{\bar m}}
I^{j'}_{c,\bar m}+\frac {B(\phi)}{2(\bar m + 1)\nu_{\bar m}}
\,.
\end{equation}
It is easy to see that the above equation gives Eq. (\ref{barsol}) for $\bar m =1$

\end{multicols}
\widetext
\begin{table}
\begin{tabular}{cccccccc}
 & $d_{Fc,0}$ & $N^h_{c,0}$& $N_{c,0}$& $d_{Fc,\gamma >0}$ & $N^h_{c,\gamma >0}$&  
$N_{c,\gamma >0}$\\
\tableline
\tableline
G.S. &$N_a-N_f$&$N_a-2N_f$&$N_f$&$N_a-2N_f$&$N_a-2N_f$&$0$\\
Ex.S.$^{\gamma}$ &$N_a-N_f+(\gamma-1)N_{c,\gamma}$&$N_a-2N_f+2\gamma 
N_{c,\gamma}$&
$N_f-(\gamma+1)N_{c,\gamma}$&$N_a-2N_f+N_{c,\gamma}$&$N_a-2N_f$&$N_{c,\gamma}$\\
\end{tabular}
\caption{Available and occupied quantum numbers $I_{c,\gamma}$
for the spinless fermion model. In first column G.S.
and Ex.S.$^{\gamma}$ stand for the ground state and for the 
excited states with $N_{c,\gamma>0}$
pseudoparticles in a single $\gamma$ band, respectively. In the first row, 
$d_{c,\gamma}$ gives the available $I_{c,\gamma}$ quantum numbers, $N^h_{c,\gamma}$
gives how many $I_{c,\gamma}$ are unoccupied, and $N_{c,\gamma}$ is the number
of $c,\gamma$ pseudoparticles. The numbers $N_a$ and $N_f$ are the number of 
lattice sites and the number of fermions, respectively.}
\label{Tba}
\end{table}


\begin{references}
\bibitem{Maldague77}
Pierre F. Maldague,
Phys. Rev. B {\bf 16}, 2437 (1977).

\bibitem{Loh88}
E. Y. Loh and D. K. Campbell,
Synth. Metals {\bf 27}, A499 (1988).

\bibitem{Schulz90}
H. J. Schulz, Phys. Rev. Lett 
{\bf 64}, 2831 (1990); H. J. Schulz, {\it Fermi Liquids and non-Fermi
Liquids}, in {Les Houches, Session} LXI, 1994, edited by E. Akkermans,
G. Montambaux, J.-L. Pichard, and J. Zinn-Justin (Elsevier, 1995). 

\bibitem{Stafford90}
 C.A. Stafford, A.J. Millis, and B.S. Shastry, Phys. Rev. B {\bf 43},
13660 (1990); 
C. A. Stafford and A. J. Millis, Phys. Rev. {\bf 48}, 1409 (1993).

\bibitem{Fye91}
R. M. Fye, M. J. Martins, D. J. Scalapino, J. Wagner, and W. Hanke,
Phys. Rev. B {\bf 44}, 6909 (1991).

\bibitem{Giamarchi92}
T. Giamarchi and A. J. Millis,
Phys. Rev. B {\bf 46}, 9325 (1992).

\bibitem{Carmelo92}
J. M. P. Carmelo and  P. Horsch,
Phys. Rev. Lett. {\bf 68}, 871 (1992);
J. M. P. Carmelo, P. Horsch, and A. A. Ovchinnikov,
Phys. Rev. B {\bf 46}, 14728 (1992).

\bibitem{Horsch93}
P. Horsch and W. Stephan
Phys. Rev. B {\bf 48}, 10595 (1993).

\bibitem{Castella95}
H. Castella, X. Zotos, and P. Prelov\v{s}ek,
Phys. Rev. Lett. {\bf 47}, 972 (1995) .

\bibitem{Castella96} H. Castella and X. Zotos
Phys. Rev. B {\bf 54}, 4375 (1996).

\bibitem{Zotos96} X. Zotos and P. Prelov\v{s}ek,
Phys. Rev. B {\bf 53}, 983 (1996).

\bibitem{Antonio96}
A. H. de Castro Neto and M. P. A. Fisher, Phys. Rev. B
{\bf 53}, 9713 (1996).

\bibitem{Naef97} X. Zotos, F. Naef, and P. Prelov\v{s}ek,
Phys. Rev. B {\bf 55}, 11029 (1997).

\bibitem{Sacramento94} P.D. Sacramento, 
Z. Phys. B {\bf 94}, 347 (1994).

\bibitem{Eggert94} Sebastian Eggert, Ian Affleck, and Minoru Takahashi,
Phys. Rev. Lett. {\bf 73}, 332 (1994).

\bibitem{Wells95} B. O. Wells, Z.-X. Shen, A. Matsuura, D.M. King, M.A.
Kastner, M. Greven, and R.,J. Birgeneau, Phys. Rev. Lett.
{\bf 74}, 964 (1995).

\bibitem{Motoyama96} N. Motoyama, H. Eisaki, S. Uchida, Phys. Rev. Lett.
{\bf 76}, 3212 (1996).

\bibitem{Kim96} C. Kim, A.Y. Matsuura, Z.-X. Shen, N. Motoyama, H. Eisaki,
S. Uchida, T. Tohyama, and S. Maekawa, Phys. Rev. Lett.
{\bf 77}, 4054 (1996).

\bibitem{Mori97} T. Mori, T. Kawamoto, J. Yamaura, T. Enoki, Y. Misaki,
T. Yamabe, H. Mori, S. Tanaka, Phys. Rev. Lett. {\bf 79}, 1702,
(1997).

\bibitem{Kohn64} W. Kohn, Phys. Rev. {133}, A171 (1964).


\bibitem{Yang66} C.N. Yang and C.P Yang, Phys. Rev. {\bf 150},
321 (1966).

\bibitem{Cloizeaux62} Jacques Des Cloizeaux and J.J. Pearson,
Phys. Rev. {\bf 128}, 2131 (1962).

\bibitem{Cloizeaux66} Jacques Des Cloizeaux and Michel Gaudin,
J. Math. Phys. {\bf 7}, 1384 (1966).

\bibitem{Lieb68} Elliott H. Lieb and F. Y. Wu,
Phys. Rev. Lett. {\bf 20}, 1445, (1968).


\bibitem{Kawakami97} Satoshi Fujimoto and Norio Kawakami, 
J. Phys. A {\bf 31}, 465 (1998).


\bibitem{Takahashi97}
For an excellent review, see Minoru Takahashi, preprint, cond-mat. 9708087.

\bibitem{Kirchner98} S. Kirchner, H. G. Evertz, and 
W. Hanke, preprint, cond-mat. 9804148.

\bibitem{Tomaz98} Toma\v{z} Prosen, Phys. Rev. Lett. {\bf 80}, 1808 (1998).

\bibitem{Shastry86} B. S. Shastry, Phys. Rev. Lett. {\bf 56}, 1529 (1986).

\bibitem{Narozhny98} B. N. Narozhny, A. J. Millis, and 
N. Andrei, preprint, cond-mat. 9711100.

\bibitem{Carm91a}
J. Carmelo and A. A. Ovchinnikov, J. Phys.: 
Condens. Matter {\bf 3}, 757 (1991).

\bibitem{Carm91b} J. Carmelo, P. Horsch, P.-A. Bares, and A. A. Ovchinnikov,
Phys. Rev. B {\bf 44}, 9967 (1991).

\bibitem{Carm98} J. M. P. Carmelo, P. Horsch, 
A. A. Ovchinnikov, D. K. Campbell, 
A. H. Castro Neto, and N. M. R. Peres, Phys. Rev. Lett. {\bf 81}, 489 (1998).

\bibitem{Baym}
See, for example, Gordon Baym and Christopher J. Pethick,
in {\em Landau Fermi-Liquid Theory 
Concepts and Applications},(John Wiley \& Sons,
New York, 1991).

\bibitem{Haldane81}
F. D. M. Haldane, J. Phys. C {\bf 14}, 2585 (1981).

\bibitem{Haldane91}
F. D. M. Haldane, Phys. Rev. Lett. {\bf 67}, 937 (1991). 

\bibitem{Gaudin72} M. Gaudin, Phys. Rev. Lett. {\bf 26}, 1301 (1971).

\bibitem{Shastry90} B. Sriram Shastry and Bill Sutherland,
Phys. Rev. Lett. {\bf 65}, 243, (1990).

\bibitem{Nuno97a} N.M.R. Peres, J.M.P. Carmelo, D.K. Campbell,
and A.W. Sandvik, Z. Phys. B
{\bf 103}, 217 (1997).

\bibitem{Nuno97b} J.M.P. Carmelo and 
N.M.R. Peres, Phys. Rev. B {\bf 56}, 3717 (1997).

\bibitem{Nuno97c} N. M. R. Peres, 
Ph. D. Thesis, cond-mat. 9802240.

\bibitem{Carmelo98p} J.M.P. Carmelo, P.D. Sacramento, N.M.R. Peres, and
D. Baeriswyl, preprint (1998).

\bibitem{Imada95} In the context of the metal-insulator 
transition driven by interactions, Imada has proposed a scaling 
theory based on the scaling behavior of the charge 
stiffness and the compressibility upon doping. He developed
an interpretation of the scaling theory of the metal-insulator 
transition in terms of a {\it band picture} for the elementary excitations, 
very much in the spirit discussed in this work for our one-dimensional system. 
See M. Imada, J. Phys. Soc. Jpn. {\bf 64}, 2954 (1995); H. Tsunetsugu and   
M. Imada, J. Phys. Soc. Jpn. {\bf 67}, 1864 (1998) and references 
therein.

\bibitem{Shankar} It is worthwhile to note that a mean-field treatment
of Hamiltonian (\ref{spinless}) also 
gives an interpretation of the metal-insulator transition as a band
insulator, although it gives results that are incorrect in an
important qualitative manner (it predicts the 
metal-insulator transition for $n=1/2$ and {\it any non-zero value} of $V$),.
See, for example, Shankar, Rev. Mod. Phys. {\bf 66}, 129 (1994).

\bibitem{Kawakami91} See, for example, a similar 
calculation for the super-symmetric $t-J$ and 
Hubbard models, Norio Kawakami and Sung-Kil Yang, 
Phys. Rev. B {\bf 44}, 7844 (1991).

\bibitem{comm1} It is interesting to remark that this result, being true for
a free and tight-binding electron systems, is somewhat unexpected in a strongly
correlated 1D system. Again, this seems to be a result of integrability and 
fits well with the band picture for the elementary excitations described above.

\bibitem{Haldane81b} The velocity vanishes linearly with the doping 
$\delta=\vert 1/2-n \vert$. F.D.M. Haldane, Phys. Lett. {\bf 81A}, 153 (1981).

\bibitem{Kusmartsev92} F. V. Kusmarstev, Phys. Lett. A {\bf 161}, 433 (1992).

\bibitem{comm} See Ref. \onlinecite{Kusmartsev92} for
the differences between the anysotropic Heisenberg
chain (``a-cyclic" problem) and the spinless fermion
model (``c-cyclic" problem).

\bibitem{comm2} Equation (\ref{energyasymp})
is equivalent to the diagonalization
of  Hamiltonian (\ref{spinless}) in the Hilbert
subspaces of one, two, ..., $n$ nearest-neighbor occupied
sites. In each of these subspaces the energy eigenvalues are
of the order of $E_0$, $E_0+V$, ..., $E_0+nV$, respectively,
where $E_0$ is the energy of the ground state.

\bibitem{comment} In a previous version of this work
we had concluded that $D_n$ was finite, 
in the half-filled band case, due to an innappropriate
perturbative treatment of the BA equations. That 
treatment prevented us to detect the elusive cancelation of the flux factors that we show in this paper.

\bibitem{Table} {\it Table of Integrals, Series, and Products},
Edited by I.S. Gradshteyn and I. 
Ryzhik, 
Academic Press, 5. ed., p. 916.

\end{references}
\end{document}